# Anisotropy of Galactic Cosmic Rays and New Discoveries in Its Measurements

Yu. M. Andreyev, V.A. Kozyarivsky and A.S. Lidvansky

Institute for Nuclear Research, Russian Academy of Sciences, Moscow

## Abstract

We discuss recently published results of two-dimensional measurements of the cosmic ray anisotropy in the energy range 1-100 $TeV$. It is demonstrated that, in spite of pretence of the authors to measure the anisotropy in more detail than it was done in one-dimensional measurements of the first harmonic of CR intensity in sidereal time, new measurements give nothing essentially new. Moreover, two-dimensional picture is misleading creating an illusion that the true direction of the anisotropy is observed, while, as before, only the projection of the anisotropy onto the equatorial plane is measured and the phase of the anisotropy remains to be the only directly measured parameter. The sophisticated interpretations of the results of 2D measurements made by their authors are invalid, since they are based on the false assumption that the equatorial excess and deficit of CR intensity seen on the difference maps represent the real anisotropy.

**Introduction**

Galactic cosmic rays (GCR) have a rather small anisotropy that is measurable in the energy range where the solar modulation of GCR stop acting (approximately above 1 *TeV*), but the counting rate at detectors of reasonable dimensions is still high enough to ensure the necessary statistical accuracy (approximately up to 100 *TeV*). In this energy range the experiments are performed either under a heavy overburden of rock (water, ice) using large-area cosmic-ray muon detectors or on the ground surface with arrays counting small-size air showers. Recently, two giant international collaborations representing muon underground detectors (Super–Kamiokande in Japan) and surface air shower arrays (Tibet ASγ in China) published their data on the anisotropy of GCR [1, 2, 3, 4]. These data have been recently supplemented by the results of the water Cherenkov detector Milagro [5] presented at the last 30th International Cosmic Ray Conference (Merida, Mexico, 2007). All these three facilities presented two-dimensional (right ascension -- declination) maps of the events, and in all cases the authors believe that the two-dimensional measurements provide for better knowledge of the real galactic anisotropy. One of the main goals of this paper is to call this belief in question. Another goal is to discredit the interpretation currently made on the basis of the data of these measurements. And, finally, we would like to present a discussion of the general problem of GCR anisotropy measurements, as it is seen by us at the moment. To this end, we first consider the case of simple dipole anisotropy.

**Dipole anisotropy of GCR and diurnal wave of intensity**

The dipole anisotropy of GCR can be represented as a vector whose magnitude is equal to $\xi$. This vector is directed to a celestial point with coordinates $\alpha_0$ , $\delta_0$ in equatorial coordinates (or $l_0$ , $b_0$ in galactic coordinates). In the first approximation, the directional distribution of cosmic ray intensity has the following form [6]:

$$I(\theta) = I_0 + i_0 \cdot Cos\,\theta\,, \qquad (1)$$

where $\theta$ is the angle between the true anisotropy direction and the direction of observation at a given time instant. Due to rotation of the Earth this angle varies with time as

$$Cos\,\theta(t) = Sin\,\delta_0 Sin\,\delta_T + Cos\,\delta_0 Cos\,\delta_T Cos(t-t_0)\,, \qquad (2)$$

where $\delta_T$ is the declination of an observing telescope (see Fig. 1, formula (2) is obtained from the spherical triangle with vertexes at the pole N and points $\alpha_0$, $\delta_0$ and $\alpha_T$, $\delta_T$). Substituting (2) into (1) we obtain the sidereal daily wave of GCR intensity as the following sum

$$I(t) = I_0 + i_0 Sin\,\delta_0 Sin\,\delta_T + i_0 Cos\,\delta_0 Cos\,\delta_T Cos(t-t_0)\,. \qquad (3)$$

After normalization to the isotropic part of intensity $I_0$ formula (3) takes on the form

$$\frac{I(t)}{I_0} = 1 + \xi Sin\,\delta_0 Sin\,\delta_T + \xi Cos\,\delta_0 Cos\,\delta_T Cos(t-t_0)\,. \qquad (4)$$

In this expression the sum of the first and second terms in the right-hand side represents zero (time-independent) harmonic. The third term is the first harmonic in sidereal time that is usually measured. Its amplitude is proportional to the degree of the anisotropy $\xi$, to cosine of telescope declination $\delta_T$, and to the cosine of the anisotropy vector declination. Analyzing the first harmonic one can derive its phase $t_0$, corresponding to right ascension $\alpha_0$, which due to this fact is the directly measured parameter of sidereal anisotropy. Formula (4) is exact for ideal narrow-angle telescope. For a wide-angle telescope it should be replaced by a similar formula with integration made taking the acceptance function into account (see below). The standard approach (see, for example, [21]) is to neglect the second term in the right-hand side of (4), assuming zero harmonic to be equal to unity. Under analysis is in this case only the first harmonic whose amplitude is determined experimentally. Upon calculation, using the acceptance of the telescope, of the effective declination, one can then consider the real wide-angle telescope as a narrow-angle telescope with declination $\delta_{eff}$ and assume cosine dependence still valid. Dividing the amplitude of the first harmonic by $Cos\,\delta_{eff}$ one reduces the data to the equator (zero declination). The result is the projection of the anisotropy vector onto the equatorial plane. It is not equal to the degree of the true anisotropy, since it includes an uncertain factor $Cos\,\delta_0$. Thus, not only one is unable to determine the anisotropy declination by analyzing the daily wave of intensity in sidereal time, but this unknown declination introduces uncertainty into the anisotropy magnitude, which is not equal to the diurnal wave amplitude even after reducing it to the equator. One should also notice that measurements of the first harmonic in northern and southern hemispheres (even using identical telescopes) cannot help one to improve the situation in principle. Measurements of identical telescopes in different hemispheres at declinations equal in absolute value would yield one and the same result, since cosine is even function.

**Results of Super-Kamiokande and Tibet ASγ Collaborations**

In papers [1, 2] the Super-Kamiokande collaboration presented the celestial map (Fig. 2) with an excess of events (0.104 ± 0.020)% from the region of Taurus constellation, the center of gravity of this excess being located at a declination of –5° ± 9° (the Virgo deficit observed on the Super-Kamiokande map is also close to the equator). A similar map (Fig. 3) was published by the Tibet ASγ collaboration [3, 4], and their excess of events is also at zero declination. Both maps have the regions of excess and deficit of events. Taking so far apart the question of possible complex structure of the anisotropy (the presence of the second harmonic etc.) let us look at the general structure of the maps. Based on the simple dipole anisotropy of formula (1) one could expect that, according to formula (4), the first harmonic is maximum at zero declination owing to the factor $Cos\delta_T$ (which represents the direction of observation in the case of a narrow-angle telescope). This is just the case in both presented maps. In actual fact the data published in [1, 2, 3, 4] measure this cosine dependence, and concentration of the excess on the equator is a trivial consequence of the structure of formula (4). It is due to this fact that usually, when measuring the first harmonic of intensity in the sidereal time, the amplitude is reduced to the equator (without this operation it is impossible to compare the data of different detectors) dividing it by the cosine of effective declination, as we discussed above.

It should be also noted that, in addition to the two-dimensional map, the Super-K collaboration presented the variation of the muon counting rate as a function of right ascension, i.e., the result of usual one-dimensional measurement. The amplitude of this wave slightly differs in different publications, we have taken the value $(5.3 \pm 1.2) \times 10^{-4}$ from [1]. This amplitude is twice lower than the Taurus excess, and this is a good demonstration of what was said above about reduction to the equator: the factor of two represents the ratio of cosines of zero declination and effective declination of the muon telescope. The phase of the first harmonic is equal to $40° \pm 14°$, which should be compared with the right ascension of the Taurus excess $75° \pm 7°$.

The fact that the Taurus excess is an artifact produced by a method of map construction can be easily seen if we analyze how this method works.

**One-dimensional and two-dimensional measurements**

The flux of particles from point $(\alpha, \delta)$ is modulated by the thickness of the atmosphere and (in the case of underground facility) also by the thickness of rock. Without loss of generality, we consider now only modulation by the atmosphere (the case of extensive air showers) as more regular (there is no need to know complicated distribution of matter in underground experiments). Let a$(A, z)$ be the counting rate in the horizontal system $A$, $z$ (azimuth and zenith angles), normalized to unity. Then for an arbitrary EAS array the modulating profile as a function of declination looks like

$$\rho(\delta) = \int_0^{2\pi} a(A(\alpha, \delta), z(\alpha, \delta)) d\alpha \ , \qquad (5)$$

since integration over $\alpha$ along the line $\delta = const$ is equivalent to the passage of corresponding point $(\alpha, \delta)$ throughout the full sidereal day. An example of this modulating profile, calculated for extensive air showers with angular distribution $dI \sim cos^7 \theta \, d\Omega$ within the cone of $40°$, is shown in Fig. 4. After rewriting formula (4) in the two-dimensional form

$$\frac{I(\alpha, \delta)}{I_0} = 1 + \xi \sin \delta_0 \sin \delta + \xi \cos \delta_0 \cos \delta \cos(\alpha - \alpha_0), \quad (6)$$

one can write the measured flux multiplying (6) by modulating profile (5):

$$I_{meas}(\alpha, \delta) = \rho(\delta)(1 + \xi \sin \delta_0 \sin \delta) + \rho(\delta)\xi \cos \delta_0 \cos \delta \cos(\alpha - \alpha_0). \quad (7)$$

We read in the Super-Kamiokande paper: "The muon flux from a given celestial position can be directly compared with the average flux for the same declination" [1]. This means that this average flux is subtracted from the flux in each cell of the given declination belt. But since the map is presented (see Fig. 2) in percentage or in standard deviations, it is obvious that normalization is made to the average value at each declination. Tibet ASγ writes about the same even more candidly: "..the average intensity in each narrow Dec belt is normalized to unity" [3]. We can easily make averaging and normalization by dividing (7) by its averaged value. Then, independent of acceptance, we have

$$I'_{meas}(\alpha, \delta) = 1 + \frac{\xi \cos \delta_0 \cos \delta}{1 + \xi \sin \delta_0 \sin \delta} \cos(\alpha - \alpha_0) \quad . \quad (8)$$

This formula exactly represents the two-dimensional maps under discussion, and if we are interested in the maximum of this distribution in $\delta$, we can easily demonstrate that this maximum lies at

$$\sin \delta_{\max} = -\xi \sin \delta_0 \quad . \quad (9)$$

The difference with formula (4) where the first harmonic should have its maximum on the equator is due to the fact that formula (8) includes zero harmonic too, which results in a small shift of the maximum. However, since $\xi$ in formula (9) is very small, $\delta_{max}$ is fairly close to zero. Nevertheless, if $\delta_0$ is positive, there should be formally a small shift of the maximum into the negative region (it is funny that Super-Kamiokande data indeed demonstrate such a shift, though too big to be explained by this effect).

Figures 5 and 6 present the world data of one-dimensional measurements to be compared with the new two-dimensional measurements. We have included in these plots only those results for which the reduction to the equator was made or possible to be made. One can see that the result of Super-K, taken either as the Taurus excess or as the first harmonic amplitude value divided by cosine of effective declination (which is approximately the same), is in a reasonable agreement with other data. But this is the case only as far as the amplitude of the effect is concerned. The phase of the first harmonic (40° ± 14°) is well consistent with world data, while the position of the Taurus excess found using some method of maximization of the effect (which seems to be statistically incorrect procedure) is $75° \pm 7°$, which is only marginally consistent with the majority of experiments. This is the only result of construction of the map and its sophisticated processing.

In Table we demonstrate the result of more direct comparison of the Super-K data with our measurements made with the Baksan Underground Scintillation Telescope (BUST) [7]. The daily wave in sidereal time measured by the BUST (see Fig. 7) is less in amplitude than that of Super-K (the latitude difference between BUST and Super-K is 13°), though measured with better accuracy.

Table

| Telescope | Super-K | BUST |
|---|---|---|
| Effective energy of primaries | 10 TeV | 2.5 TeV |
| Amplitude of first harmonic | $(5.3 \pm 1.2) \times 10^{-4}$ | $(3.79 \pm 0.27) \times 10^{-4}$ |
| Phase of the first harmonic | $2.67 \pm 0.93$ h RA | $1.84 \pm 0.28$ h RA |
| Projection onto the equator | $(10.4 \pm 2.0) \times 10^{-4}$ (Taurus excess) | $(10.11 \pm 0.72) \times 10^{-4}$ (divided by $Cos\delta_{eff}$) |
| Statistical significance | $5\sigma$ | $14\sigma$ |

In order to reduce it to the equator, we divided it by cosine of the effective declination (calculated value of the effective declination is 68°). This projection of the true anisotropy on the equator almost coincides with the Taurus excess (see Table). The phases of the first harmonics in these two experiments are in good agreement. So, the one-dimensional experiment performed by Super-Kamiokande collaboration yielded a reasonable result consistent with others. The two-dimensional analysis made by the same collaboration produced nothing new except for the wrong interpretation and a shift of the anisotropy phase by more than 3 h in right ascension.

**Monte Carlo analysis of two-dimensional measurements**

The arguments presented above seem to be very simple if not trivial. Nevertheless, for reasons unknown to us, it turned out that it was difficult for some people to catch the point of controversy, as showed our discussions with many people especially during the 30th International Cosmic Ray Conference in Merida, Mexico. So, we have made a simple Monte Carlo calculation reproducing the method of map construction used by the Super-Kamiokande and Tibet ASγ collaborations. Figure 9 presents the map of ideal dipole anisotropy as if measured with infinite accuracy beyond the Earth. Parameters of this anisotropy are arbitrary, though its degree is taken equal to the value estimated by us in [10] as the true anisotropy amplitude, $\xi = 0.2\%$ (this value is connected with $\delta_0 = 60°$). The maximum is assumed to be at point $\alpha_0 = 14$ h RA, $\delta_0 = 60°$. The real dipole anisotropy (with statistical accuracy of measurement taken into account), detected without terrestrial effects, is shown in Fig. 10. Simulation is made according to the normal law in each cell with dimensions 2°x2°, assuming a mean value of $10^6$ and a standard deviation of $10^3$. This distribution should be multiplied by the acceptance factor of each cell, uniform in the right ascension and distributed in declination as the modulating profile of Fig. 4. The result is shown in Fig. 11, and it is this distribution with which a researcher should work as experimental data. We applied to the map of Fig. 11 the same procedure as Super-Kamiokande and Tibet ASγ collaborations: averaging in narrow declination bands with subsequent subtraction of the averaged value in each cell and normalization. The result of this procedure is seen in Fig. 12: both maximum and minimum of the map lie near the equator (zero declination), though we put into simulation the anisotropy with $\delta_0 = 60°$.

**Discussion and Conclusions**

In paper [2] many formulas of Section A coincide with ours. However, formula (4) of this Section used for the map construction includes averaging procedure, as a result of which the maximum of obtained two-dimensional distribution appears at zero declination, independent of the real anisotropy maximum $\delta_0$. This is evident from our formulas (8) and (9) for the dipole anisotropy. Formulas (12) and (13) of paper [2] also have the maximum at zero declination, though the remark of the authors that (13) "is the form of the anisotropy for the projection of the original dipole in the equatorial plane" is quite true.

What if the real anisotropy is more complicated than in our model, and its structure is not dipole? The Tibet ASγ collaboration especially insists [3] on the peculiar anisotropy structure, existence of the second harmonic, etc. Some people disputed our arguments on this ground. In this case, the complex anisotropy function can be expanded in terms of simpler functions, for example, Legendre polynomials. Any (doubly differentiated and real) function $\Phi(\delta, \alpha)$ $(-\pi/2 \leq \delta \leq \pi/2,\ 0 \leq \alpha \leq 2\pi)$ defined on the spherical surface admits expansion into absolutely and uniformly converging series [22]:

$$\Phi(\delta,\alpha) = \sum_{j=0}^{\infty}\left[\tfrac{1}{2}a_{j0}P_j(\sin\delta) + \sum_{m=1}^{j}P_j^m(\sin\delta)(a_{jm}\cos(m\alpha) + b_{jm}\sin(m\alpha))\right], \quad (10)$$

where constants $a_{j0}$, $a_{jm}$, and $b_{jm}$ are determined by function $\Phi(\delta, \alpha)$, $P_j(\sin\delta)$ are Legendre polynomials, and $P_j^m(\sin\delta)$ are associated Legendre polynomials. Retaining in this sum only the constant term and the first harmonics in angles $\delta$ and $\alpha$ we get in such an approximation:

$$\Phi(\delta,\alpha) \approx A_0 + A_1\sin\delta + \cos\delta(A_2\cos\alpha + A_3\sin\alpha). \quad (11)$$

If one sets

$$A_{23} = \sqrt{A_2^2 + A_3^2}, \quad \cos\alpha_0 = A_2/A_{23}, \quad \sin\alpha_0 = A_3/A_{23},$$

$$A_4 = \sqrt{A_1^2 + A_{23}^2}, \quad \cos\delta_0 = A_{23}/A_4, \quad \sin\delta_0 = A_1/A_4,$$

then

$$\Phi(\delta,\alpha) \approx A_0 + A_4\left[\sin\delta_0\sin\delta + \cos\delta_0\cos\delta\cos(\alpha - \alpha_0)\right]. \quad (12)$$

It is obvious that the structure of formulas (4) and (12) is essentially the same. Thus, all arguments said above remain valid for an arbitrary form of the anisotropy distribution. By the way, the fact that maxima and minima of two-dimensional maps constructed by the Super-Kamiokande and Tibet ASγ collaborations lie on the equator is an indicator of good quality of these data: any deviation from this pattern would mean that something is wrong in the processing procedure.

Can it be that the true anisotropy is oriented in accordance with the Taurus excess or at least has zero declination? Yes, it can be if it happened so by chance. The point is that this does not follow from the data, because, as we have discussed above, one can yield from them only the phase (right ascension) of the anisotropy. Its declination and amplitude cannot be deduced from the two-dimensional difference map (8) in the same manner, as they cannot be determined from the one-dimensional RA distribution. One can rather try to use originally observed two-dimensional distribution (7). If one could calculate $\rho(\delta)$ with sufficient accuracy, then, dividing distribution (7) by this function, one would be able to reconstruct the true distribution of cosmic ray intensity. Unfortunately, this method of determination of $\delta_0$ from the two-dimensional map requires high-precision calculation of function $\rho(\delta)$, which is very difficult practically (and totally hopeless for underground telescopes). Another approach was used by us in [23] in order to construct a celestial map for air showers detected in the Baksan gamma-ray astronomy experiment. The empirical angular acceptance was folded in to the exposure distribution in sidereal time to predict the two-dimensional distribution of showers for an isotropic incident flux. This expected distribution was subtracted from the experimental two-dimensional distribution. The resulting map in standard deviations looks more or less reasonable from the point of view of our present-day knowledge[1]: the map, constructed using the method differing from that described above, demonstrated a broad maximum with the center of gravity approximately at $\alpha_0 \sim$ 2-3 h RA and $\delta_0 \sim$ 45-50°.

Nevertheless, the very fact that the construction of two-dimensional maps of any kind has some advantages is doubtful. The two-dimensional measurements distribute available statistics in many cells. When one deals with so small effects as the anisotropy under discussion, may be more reasonable strategy is to concentrate this statistics by enlarging the domains of this map, constructing in the limit two cross telescopes. In paper [8] we suggested a method of reconstructing real anisotropy declination $\delta_0$ with cross telescopes, and in [9] this method was tried giving evidence that the anisotropy vector lies in the galactic plane ($\delta_0 \sim$ 60°).

---

[1] There is a statement in paper [1] that "the first-ever celestial map of cosmic rays (> 10 TeV) was obtained" by the Super-Kamiokande collaboration. As a matter of fact, we published this celestial map obtained with the Baksan air shower array (100 TeV) seventeen years ago.

Finally, we can draw the following conclusions.

1. Existing one-dimensional experimental data are reasonably consistent, and the addition of the results of two-dimensional experiments makes no radical improvement of the situation.

2. The belief of the authors of two-dimensional experiments that they observe real anisotropy as positions of the maximum and minimum in the presented declination-right ascension maps is but an illusion produced by a certain misunderstanding concerning the procedure of map construction. (There are no doubts in this interpretation made by both collaborations, and it is clearly seen from the following facts: the Super-K team calls the map's maximum and minimum according to constellations where they lie, and both Super-K and Tibet AS-γ transform their maps into galactic coordinates in order to demonstrate the direction of cosmic ray anisotropy in the Galaxy).

3. Thus, there is no advantage in performing two-dimensional measurements instead of traditional one-dimensional analysis. Of course, all experiments discussed above are multi-purpose, and the analysis of cosmic ray anisotropy is a sort of by-product experiment lying far from research of higher priority. Nevertheless, even in this case one should be accurate with data analysis and its interpretation.

One more remark of general character can be made. The results of measurements in the range from 1 to 100 TeV show a certain tendency to a decrease of the amplitude of the diurnal wave with energy (Fig. 5). This could be associated with the fact that two groups of data (muon telescopes and air shower arrays) are concentrated (though overlapping) at different ends of this energy interval: on the average, muon detectors have lower energies. Since muon detectors are sensitive to the primary cosmic ray spectrum per nucleon, while air shower arrays – to the spectrum per nucleus, these two types of facilities deal with differing charge composition of GCR. Diffusion in the interstellar medium must depend on charge, and so, the measured cosmic ray anisotropy should be somewhat different in these cases. This difference, probably, is not so large (so far it is comparable with experimental scatter of data), but precisely because of the fact that it should exist we basically compared above the results of two detectors of one type (Super-Kamiokande and BUST).

**Acknowledgments**

The work was partially supported by the RF State Program of Support for Leading Scientific Schools, grant NSh-321.2008.2.

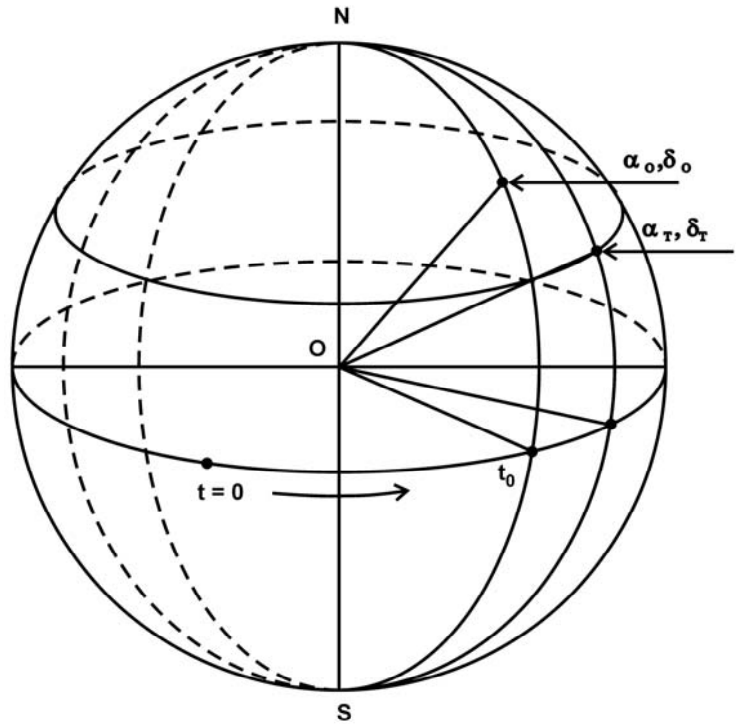

Fig.1. The true anisotropy ($\alpha_0$, $\delta_0$) and telescope ($\alpha_T$, $\delta_T$) directions in space. Formula (2) in text represents the solution of a spherical triangle with vertexes at points N, ($\alpha_0$, $\delta_0$), and ($\alpha_T$, $\delta_T$).

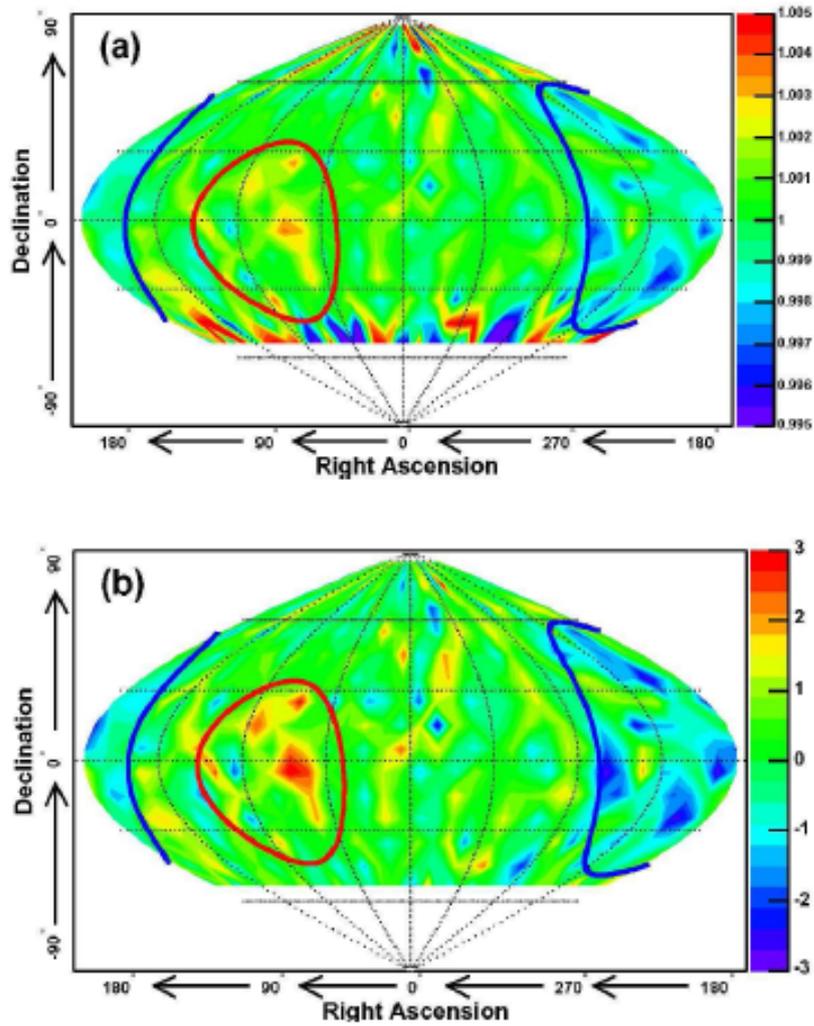

Рис 2. Primary cosmic-ray flux in the celestial coordinates as measured by Super-Kamiokande (SK) large imaging water Cherenkov detector located at 2400 m.w.e. underground in the Kamioka mine, Japan. Deviations from the average value for the same declinations are shown. The units are(a) amplitude (from -0.5% to +0.5%) and (b) significance (from -3σ to +3σ). The Taurus excess is shown by the red solid line and the Virgo deficit is shown by the blue solid line.

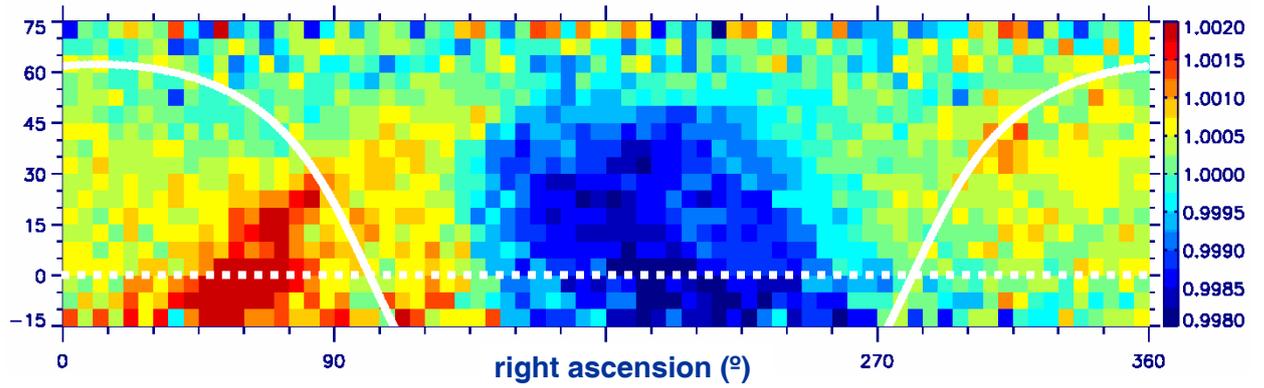

Fig. 3. GCR intensity in 5°×5° pixel measured in the Tibet AS-γ experiment and represented in a color coded format as a function of the right ascension and declination. The broken line shows the celestial equator, while the white represents the Galactic equator. The array has the modal GCR energy of ∼ 5 TeV.

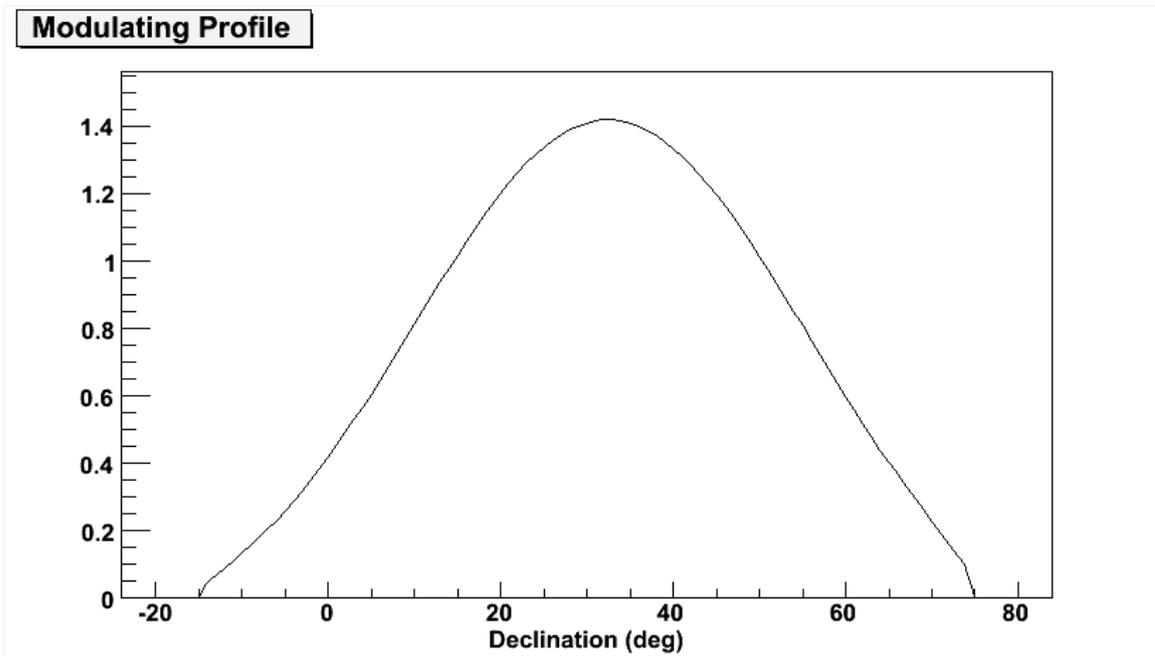

Fig. 4. Modulating profile of an EAS array (acceptance integrated over the right ascension).

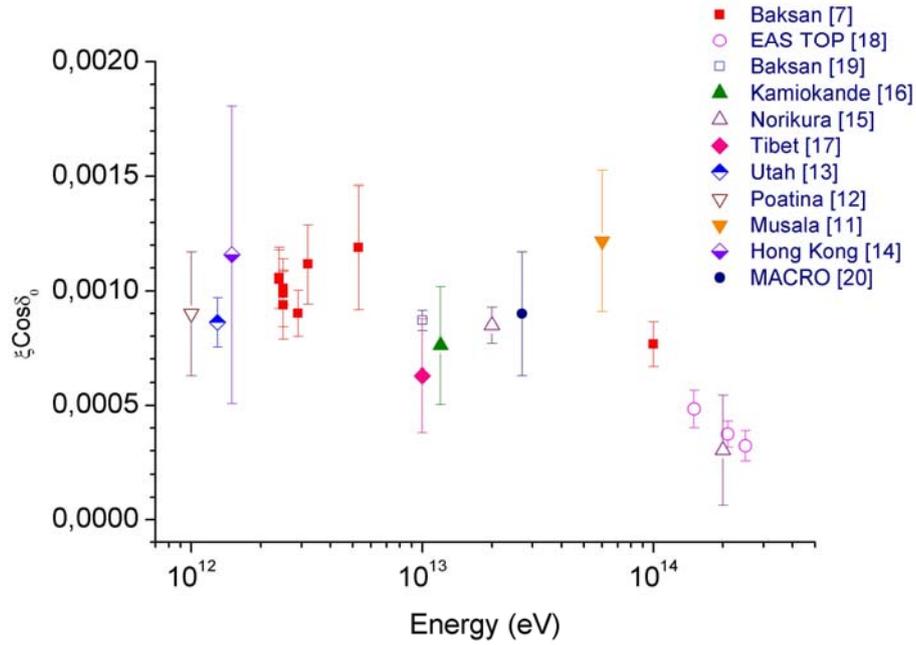

Fig 5. A summary of world data on cosmic ray anisotropy measured as the amplitude of the first harmonic in sidereal time and reduced to the equatorial plane. Only those results are presented for which this reduction was made or possible to be made. The reference number is shown after the name of each research team.

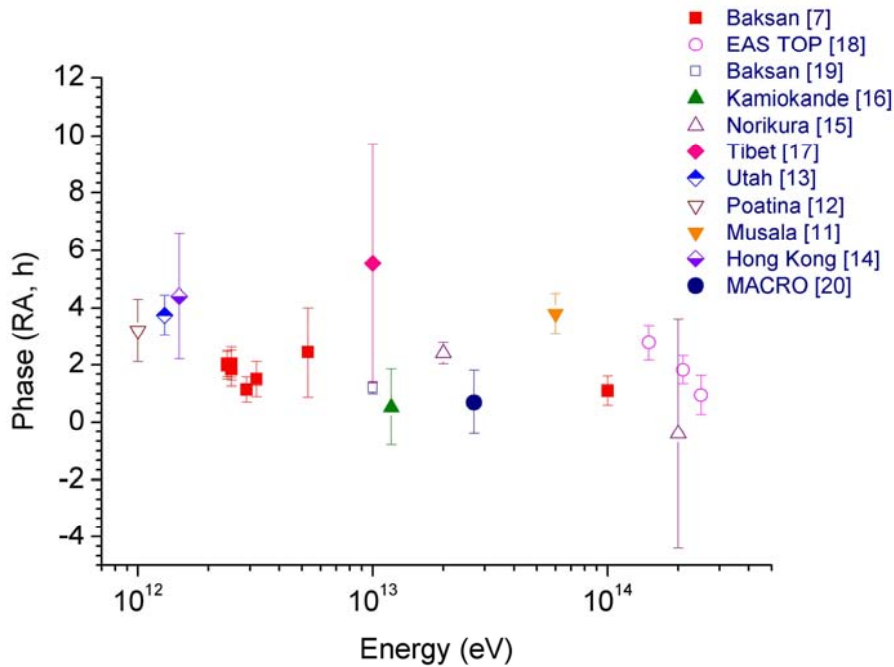

Fig. 6. Phases of the measurements of the first harmonic of intensity presented in Fig. 5.

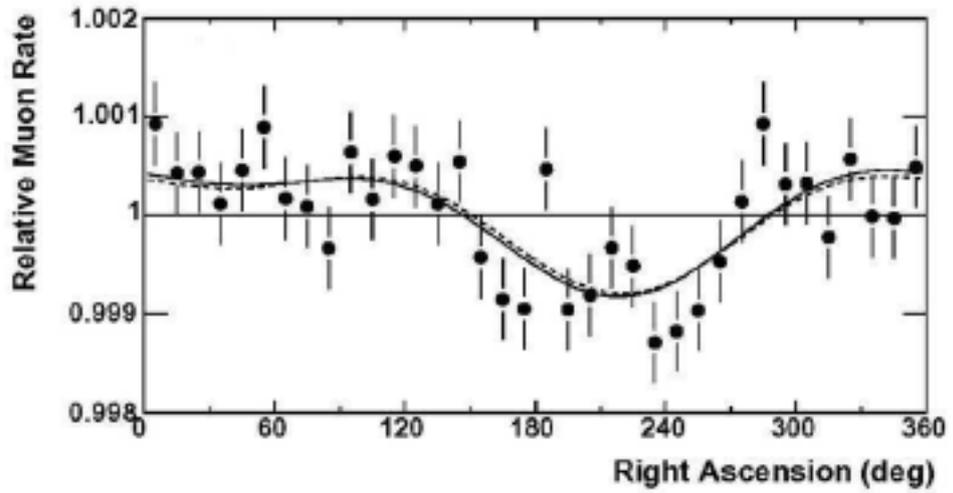

Fig. 7. The sidereal daily wave of muons in Super-Kamiokande.

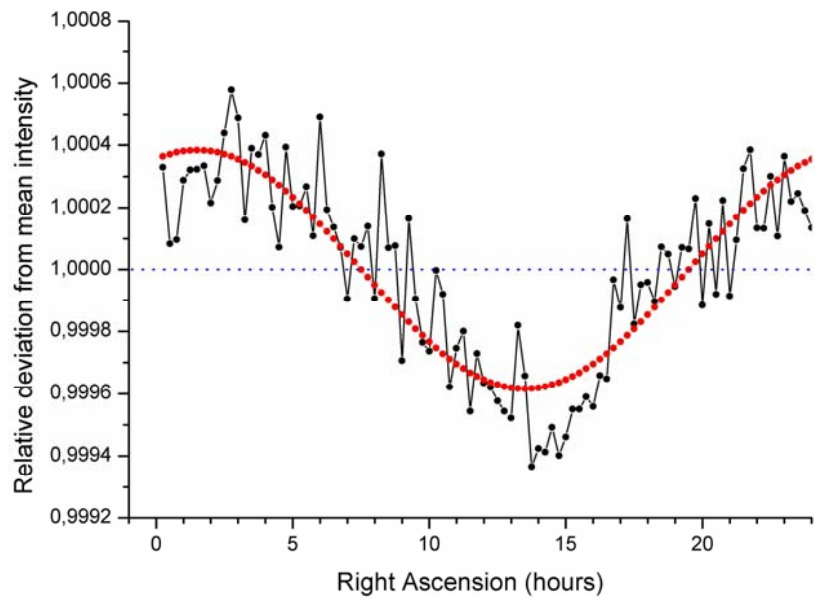

Fig. 8. The sidereal daily wave of muons as measured by the BUST in 1982-1998. The total number of events is $5 \cdot 10^9$. The threshold energy of muons 220 GeV corresponds to the energy of primary cosmic rays 2.5 TeV.

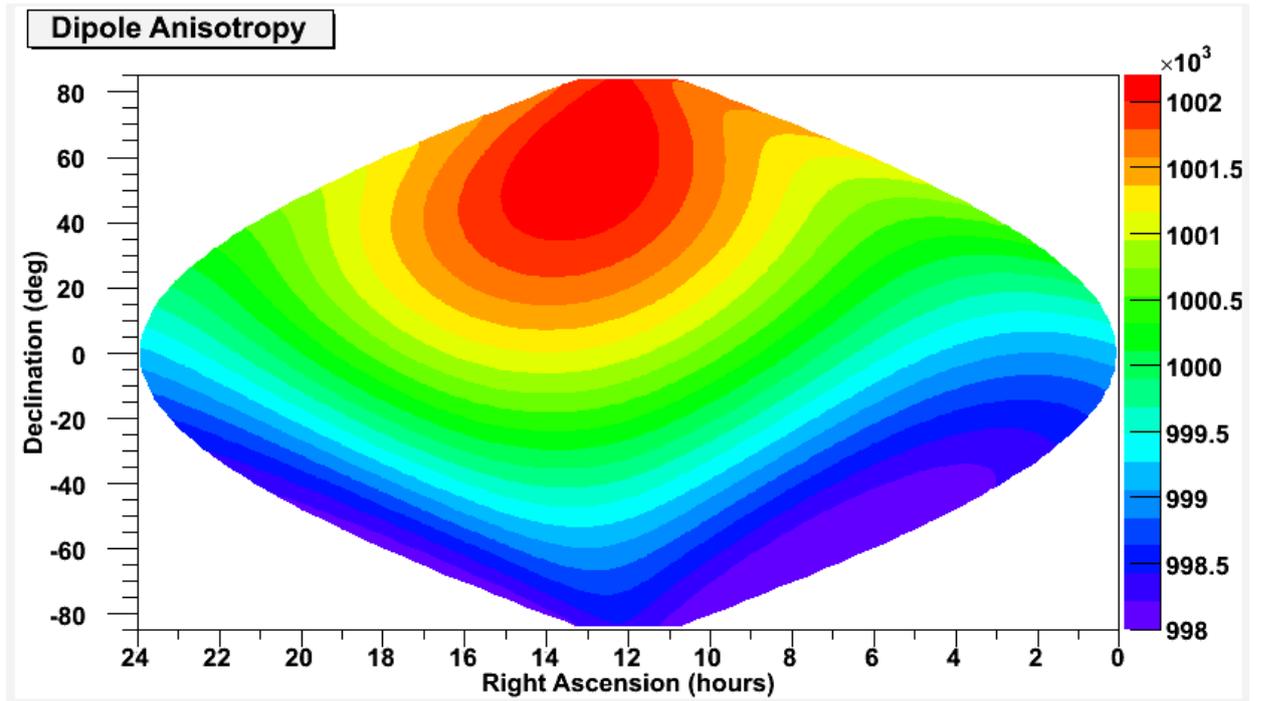

Fig. 9. The original dipole anisotropy (ideal) used for demonstration in the Monte Carlo analysis. Its distribution has maximum at point $\alpha_0 = 14$ h RA, $\delta_0 = 60°$. The degree of anisotropy is $\xi = 0.2\%$.

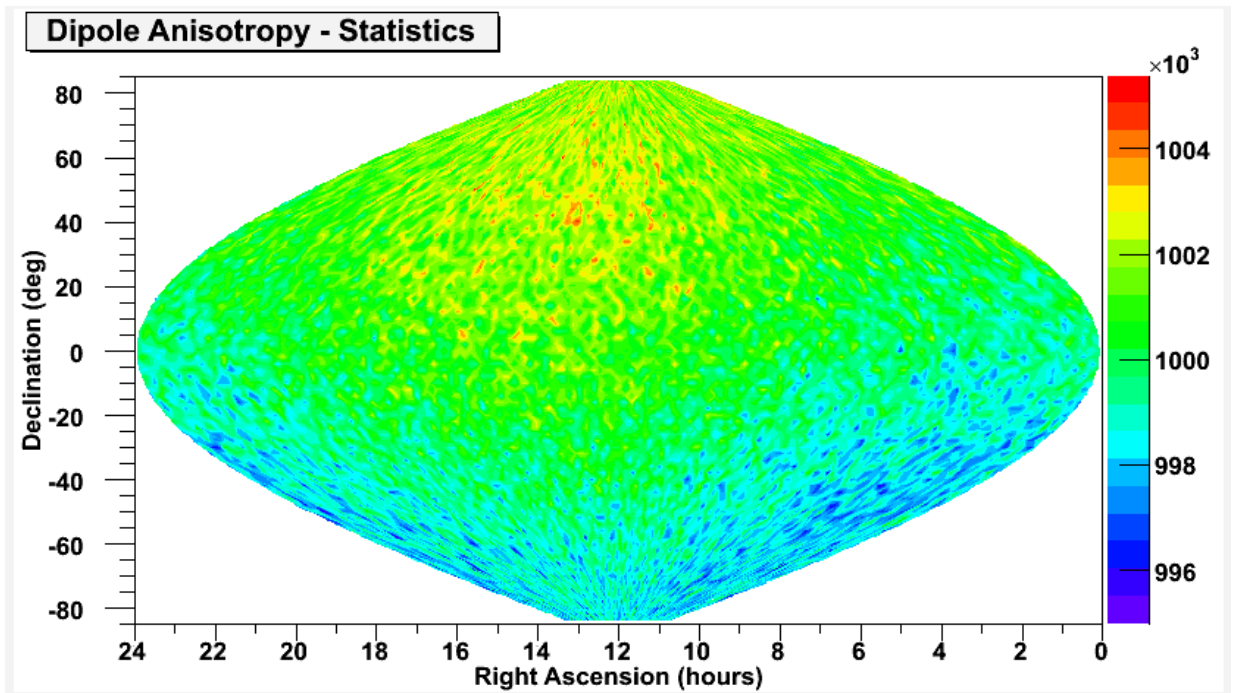

Fig. 10. The real dipole anisotropy (statistics is taken into account) detected without terrestrial effects. Simulation is made according to the normal law in each cell with dimensions 2°x2° with a mean value of $10^6$ and variance $10^3$.

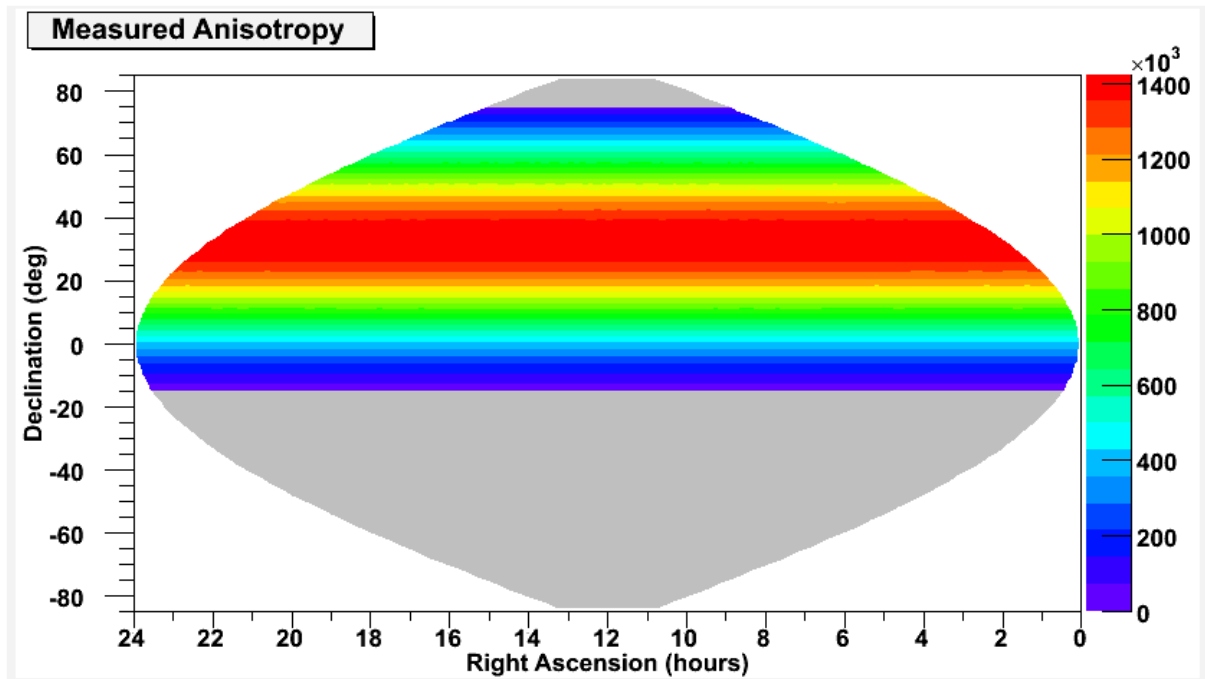

Fig. 11. The map of events detected by an EAS array on the Earth's surface. Anisotropy is included, but is not seen due to a large range of the modulating factor (see Fig. 4). The celestial region unobservable by the array is shown in gray color.

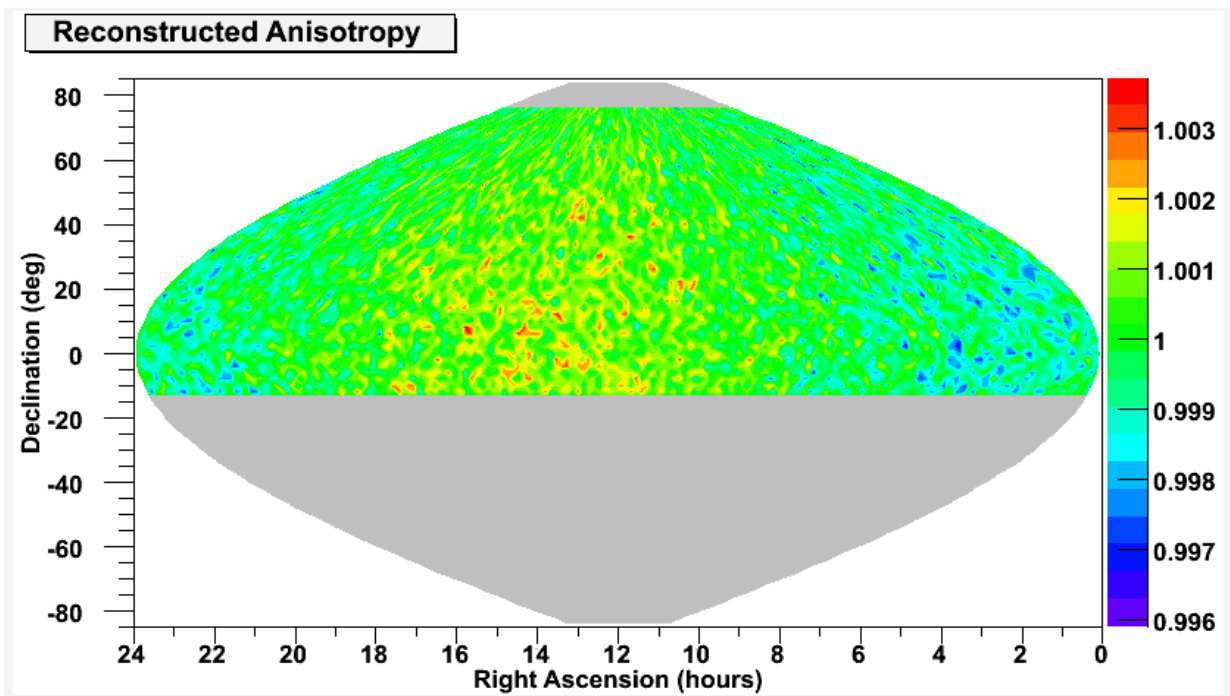

Fig. 12. The difference map after averaging in narrow declination bands and subtraction of the averaged value in every cell (this procedure reproduces the method used by Super-K and Tibet ASγ collaborations). One can see that both maximum and minimum on this map lie near the equator (though original anisotropy was quite different, see Figs. 9 and 10).